\documentstyle[11pt,psfig,twoside]{article}
\begin{document}\hbadness=10000
\thispagestyle{empty}\pagestyle{myheadings}
\markboth{J. Letessier and  J. Rafelski}{
QGP in Pb--Pb 158 A GeV collisions}
\title{Quark-Gluon Plasma in Pb--Pb 158 A GeV collisions: \\ Evidence 
from strange particle abundances and the Coulomb effect}
\author{$\ $\\
\bf Jean Letessier and Johann Rafelski\\ $\ $\\
Laboratoire de Physique Th\'eorique et Hautes Energies\\
Universit\'e Paris 7, 2 place Jussieu, F--75251 Cedex 05.\\
Department of Physics, University of Arizona, Tucson, AZ 85721\\
}
\date{}
\maketitle
\begin{abstract}
The  hadronic particle production data from relativistic 
nuclear Pb--Pb 158 A GeV collisions are
successfully described within the chemical non-equilibrium model,
provided that the analysis treats  $\Omega$ and 
$\overline\Omega$ abundances with care. We further show that there 
is a  subtle influence of the Coulomb potential on strange quarks
in quark matter which is also seen in our data analysis, 
and this Coulomb  effect confirms the finding made by chemical 
analysis in the  S--Au/W/Pb 200 A GeV collisions  that the  hadron 
particle source is deconfined with respect to strange quark propagation.
Physical freeze-out conditions (pressure, specific energy, entropy,
and strangeness) are evaluated and considerable universality 
of hadron freeze-out between the two different collision systems 
is established.\\

\noindent PACS numbers 25.75.-q, 12.38.Mh, 24.85.+p
\end{abstract}
%
\section{Introduction}
Intense experimental and theoretical work proceeds to explore
the mechanisms of quark confinement effect and the
 properties of the vacuum state
of quantum-chromodynamics (QCD), the non-Abelian gauge theory of `color'
charges \cite{FGL73}. Relativistic energy nuclear collisions are the
novel experimental tool developed in the past decade to form,
study, and explore the `melted'  space-time domain,
 where we hope to find, beyond the Hagedorn
temperature $T_{\rm H}\simeq 160$ MeV \cite{HAG},  freely
propagating quarks and gluons in the (color charge) plasma (QGP).
Analysis of hadronic particle production in relativistic 
nuclear collisions 
offers an opportunity to explore the mechanisms of 
quark confinement at the time of final state QGP freeze-out.
We present here a progress report of  such ongoing analysis of the recent and 
forthcoming experimental study of the Pb--Pb collisions at 158 A GeV 
\cite{Acta97}, which follows on our now complete analysis of the S--Au/W/Pb
collision systems \cite{LR98}.

There is little doubt that in the early Universe QGP was the transient 
state of matter, and that only about 20--40\,$\mu$sec into evolution
did our present confining vacuum freeze-out from the primordial QGP-form.
The issue is if in laboratory  experiments we can
indeed form and study the primordial QGP phase \cite{HM96}. 
In some aspects,  such as specific entropy and baryon
content, notable differences between the laboratory  QGP state 
and the early Universe conditions are present. Also,  the small,
nuclear size of the nuclear collision `micro'-bang implies a short 
laboratory lifespan $\tau_q\simeq0.5\cdot 10^{-22}$sec.  
There is also the difficult problem of proving 
the fundamental paradigm beyond a shade of doubt: can there indeed 
exist a locally  deconfined space-time domain 
with energy density exceeding by an order of magnitude
that of nuclear matter? Much of the current effort is
solely addressing this question. Among several proposed approaches to
search for and study the deconfinement, 
our work relies on the idea of strangeness
flavor enhancement, and the associated enhancement of (strange)
antibaryon formation \cite{SAB}. This signature
can be combined, in the present analysis, with the study of 
global particle abundance which represents the entropy contents of
the deconfined phase \cite{Let93,entro}.

We address here 15 presently available particle yield ratios obtained in 
central Pb--Pb 158 A GeV collision experiments carried out at CERN-SPS.
Our analysis addresses results of experiments NA49 \cite{NA49,BGS98}, 
and WA97 \cite{WA97}, we have not used results from NA44 \cite{Kan97}, 
being uncertain about the impact of the cascading weak decay
contamination of the hadronic ratios, which are quite significant.
Four WA97 data points involve $\Omega$ and $\overline\Omega$ particles
and even a cursory study of these abundances suggests  that  these 
entirely strange  particles are not falling into the same systematic
 class, a fact also visible in their unusual spectral slopes. 
We believe in view of the difference in systematics that it is appropriate 
also to consider the data excluding the  $\Omega$ and $\overline\Omega$ 
yields from analysis. In that case our analysis contains 11 relative 
experimental particle yields. However, 4 of these ratios originating 
in the same experiment are  related by a simple algebraic constraint
({\it e.g.,} $\overline\Lambda/\Lambda=\overline\Lambda/\overline\Xi \cdot
\overline\Xi /\Xi \cdot \Xi/\Lambda$), leaving us with ten independent
measurements. As we shall see, 
there are up to 5 parameters in our description. In the different 
analysis discussed here we thus have no less than 5 independent 
degrees of freedom: $ n_{\rm dof}\ge 5$. 
Since we address ratios of strange particles as well as ratios of 
total abundances of positive and negative hadrons, we combine in the present
analysis strangeness observables with the entropy enhancement \cite{Let93}.
Underlying our data analysis is the assumption of 
local thermal ({\it i.e.,} energy
equipartition) equilibrium. Both, the thermal appearance of produced
particle spectra \cite{HAG,LR98,HAGBB}, and the
qualitative and systematic agreement over many orders of magnitude
between properties of a thermal hadron system and the experimental hadron
abundance yields \cite{BSWX96}, provide a solid foundation for the 
assumption of (near) thermal equilibrium in the dense hadronic matter.

In our most recent theoretical approach there is a key
refinement not present in  earlier  work \cite{LR98}:
we do not assume chemical equilibrium even for light  quarks. 
Consideration of non-equilibrium chemical abundance for strange quarks 
allowed to analyze accurately the experimental particle
abundance data  and to characterize
 precisely the properties of the presumably deconfined
source \cite{Acta97,Let93,BGS98,Kan97,Raf91,SCRS93,Hei94,Sol97,Sae98,GS98}.  
The mechanisms of chemical equilibration
requiring reactions which change particle abundances
are today much better understood theoretically than those responsible for
what is believed to be much faster thermal (kinetic) equilibration, 
where momentum exchange between existent particles is the key mechanism.
It is hard to understand, why we, along with others, have maintained in the
past in our analysis the point of view that only strangeness has the opportunity to 
be off-equilibrium in chemical abundance. Indeed, if QGP is the particle source
the need to assimilate by fragmentation the gluon content must generate excess
light quark abundance. Since we did not allow for this freedom in earlier data 
analysis, the data were (equally badly) also  described by a high temperature 
source model~\cite{Acta97}. This is  not the case once 
also for light quarks the chemical nonequilibrium is introduced, only
the low temperature freeze-out alternative has a convincing statistical 
significance

Hadronic particles we observe are either emitted directly or are descendants
of other hadronic primaries produced near or at surface of the dense matter
fireball. Local rest-frame temperature $T$ and local collective
flow velocity $\vec v_{\rm c}$ characterize the  momentum space
distribution of particles emerging from the surface region of the fireball.
In addition, each surface volume element is characterized  by chemical
abundance factors we shall discuss in more detail below. Only
particles of similar mass and cross section experience
similar drag  forces arising from local flow of matter and hence
ratio of their abundances in some limited region of phase space
for not too small momenta 
is expected to remain unaltered by $\vec v_{\rm c}$. Since the
surface vector flow $\vec v_{\rm c}$ is a priori largely unknown, in
order to use limited `windows' of particle momenta (rapidity $y$
and transverse mass $m_\bot$) for determination of chemical freeze-out
properties of the source, only ratios of `mutually compatible' particles
can be considered, aside of  ratios of total particle
abundances. Our recent study of the flow effect shows that this procedure is
adequate, though with the ongoing improvement of the data sample we 
will soon have to include this additional freeze-out flow velocity 
parameter explicitly in the analysis~\cite{LR99}.

In the following section, we shall briefly summarize the theoretical foundations of 
the current analysis and discuss the impact of Coulomb effect in QGP on the expected
values of statistical parameters, and we shall describe the parameters that we
obtain. In section \ref{particles}, we will discuss the particle abundances and 
other related  results, such as the physical properties of the 
freeze-out dense matter. We conclude with a few brief remarks about the relevance
of our analysis for the search for QGP.

\section{Statistical model and Coulomb effect}\label{Coulomb}
The thermal production yield $dN_i$ of particles emitted within the time $dt$
from a locally at rest surface element $dS$ is:
\begin{equation}\label{Ni}
dN_i=\frac{dSd^3p}{(2\pi)^3} A_iv_idt  \,.
\end{equation}
Here $v_i=dz/dt$ is the particle velocity normal to the
surface element $dS=dx\,dy$. For a thermal  quark-gluon gas source and allowing
for recombination-fragmentation of constituents and detailed balance, the
complete phase space occupancy factor $A_i$ is given by:
\begin{equation}\label{Ai}
A_i=g_i \lambda_i \gamma_i e^{-E_i/T}\,,\qquad
\lambda_i=\prod_{j\in i}\lambda_j\,,\quad
\gamma_i=\prod_{j\in i}\gamma_j\,,\quad E_i=\sum_{j\in i} E_j \,,
\end{equation}
where $g_i$ is the degeneracy of the produced particle, and $E_i$
its energy. The valance quark content  $\{j\}$ in hadron $\{i\}$
is implied in Eq.\,(\ref{Ai}). The fugacities $\lambda_j$ arise
from conservation laws, in our context, of
quark (baryon) number and strangeness in the particle source.
$\lambda_q\equiv e^{\mu_q/T}$ is thus the fugacity
of the valance light quarks. For a nucleon $\lambda_N=\lambda_q^3$,
and  hence the baryochemical potential is: $\mu_b=3\mu_q$\,.
Similarly, for strange quarks we
have $\lambda_s\equiv e^{\mu_s/T}$\,. For an antiparticle fugacity
$\lambda_{\bar i}=\lambda_i^{-1}$\,. Some papers refer in this
context to hyper-charge  fugacity $\lambda_{\rm S}=\lambda_q/\lambda_s$,
thus $\mu_{\rm S}=\mu_q-\mu_s$.
This is a highly inconvenient historical definition arising from
considerations of a hypothetical hadron gas phase. It hides from
view important symmetries, such as $\lambda_s\to 1$ for a state in which
the phase space size for strange and anti-strange quarks is the same: at
finite baryon density the number of hyperons is always greater than the
number of anti-hyperons and thus the requirement
$\langle N_s-N_{\bar s}\rangle=0$ can only be satisfied for some
nontrivial $\lambda_s(\lambda_q)\ne 1$. Thus even a small deviation from
$\lambda_s\to1$ limit must be fully understood  in order to argue that the source
is deconfined. Conversely, observation of $\lambda_s\simeq 1$ consistently
at different experimental conditions is a strong and convincing argument
that at least the strange quarks are unbound, {\it i.e.,} deconfined.

As this discussion also illustrates, the parameter $\lambda_s$
does not regulate the total number of $s$-$\bar s$ quark-pairs present in
the system. More generally, any compound object comprising a
particle-antiparticle pair is not controlled in abundance by a fugacity,
since the formation of such particles does not impact the
conservation laws. In consequence, the abundance of, {\it e.g.,} neutral
pions comprises no quark fugacity. This is the reason to introduce an additional 
chemical phase space occupancy factor $\gamma_i$: the effective fugacity of
quarks is  $\lambda_i\gamma_i$ and antiquarks $\lambda_i^{-1}\gamma_i$. This
parameter allows to control pair abundance independently of other 
properties of the system, and in particular
temperature. For $\gamma_i\to1$ one reaches a entropy maximum \cite{entro},
corresponding to the `absolute' chemical equilibrium \cite{RM82}. Therefore
the factor  $\gamma_i$ is called the (chemical) phase space occupancy factor.

A time dependent build up of chemical abundance was first considered
in the context of microscopic strangeness production in 
QGP \cite{RM82,BZ82}, after it was realized that 
strange flavor production occurs at
the same time scale as the collision process. More generally,
one must expect, considering the time scales, that all quark flavors
will not be able to exactly follow  the rapid evolution in
time of dense hadronic matter. Moreover, fragmentation of gluons
in hadronizing QGP can contribute additional quark pair
abundance, conveniently described by the factor $\gamma_i$. It is
thus to be expected that also  for light quarks
the chemical phase space occupancy factor $\gamma_q\ne 1$.
Introduction of the factor $\gamma_q$ leads to
a  precise chemical description of the S--Au/W/Pb 200 A GeV
collisions \cite{LR98}, which was not possible before. 
The tacit choice $\gamma_q= 1$ has  not  allowed previously
to distinguish the different reaction scenarios in Pb--Pb
collisions \cite{Acta97}, where we found analyzing the experimental
data that hadronic particles could be born either at high temperature
$T\simeq 300$\,MeV or at expected hadronization temperature
$T\simeq 150$\,MeV. Introduction of $\gamma_q$, along with
 improvement in precision, allowance for quantum (Bose/Fermi) 
corrections to the Boltzmann distribution functions,
 and a greater data sample, 
allowed us moreover to recognize the systematic difference between
data points containing, and resp., not containing 
$\Omega,\,\overline\Omega$, allowing us to develop the 
precise analysis here presented.

In another refinement both $u,\,d$-flavor fugacities $\lambda_u$
and $\lambda_d$ can be introduced, allowing for up-down-quark
asymmetry \cite{Let93}. We recall that by definition $2\mu_q=\mu_d+\mu_u$,
thus $\lambda_q\equiv\sqrt{\lambda_u\lambda_d}$. For the highly 
Coulomb-charged fireballs formed in Pb--Pb collisions
a further effect of the same relative magnitude which needs consideration
is  the distortion of the particle phase space by the Coulomb potential. This
effect influences  particles and antiparticles in opposite way, and has
by factor two different strength for $u$-quark (charge $+2/3|e|$) and
($d,s$)-quarks (charge $-1/3|e|$).  Because Coulomb-effect acts in
opposite way on $u$ and $d$ quarks, its net impact on $\lambda_q$ is
relatively small as we shall see. 

However, the Coulomb effect 
distorts significantly the expectation regarding $\lambda_s\to 1$
for strangeness-deconfined source with vanishing net strangeness.
The difference between strange and anti-strange quark numbers
(net strangeness) allowing for a Coulomb
potential within a relativistic Thomas-Fermi phase space
occupancy model \cite{MR75}, allowing for finite temperature in QGP is:
\begin{equation}\label{Nsls}
\langle N_s-N_{\bar s}\rangle =\int_{R_{\rm f}} g_s\frac{d^3rd^3p}{(2\pi)^3}\left[
 \frac1{1+\gamma_s^{-1}\lambda_s^{-1}e^{(E(p)-\frac13 V(r))/T}}-
 \frac1{1+\gamma_s^{-1}\lambda_se^{(E(p)+\frac13 V(r))/T}}\right]\,,
\end{equation}
which clearly cannot vanish for $V\ne 0$ in the limit $\lambda_s\to1$.
In Eq.\,(\ref{Nsls}) the subscript ${R_{\rm f}}$ on the spatial integral 
reminds us that only the classically
allowed region within the fireball is covered in the integration over the 
level density; $E=\sqrt{m^2+\vec p^{\,2}}$, and for a uniform charge distribution
within a radius $R_{\rm f}$ of charge $Z_{\rm f}$:
\begin{equation}
V=\left\{
\begin{array}{ll}
-\frac32 \frac{Z_{\rm f}e^2}{R_{\rm f}}
      \left[1-\frac13\left(\frac r{R_{\rm f}}\right)^2\right]\,,
          & \mbox{for}\quad r<R_{\rm f}\,;\\
 & \\
-\frac{Z_{\rm f}e^2}{r}\,,& \mbox{for} \quad r>R_{\rm f}\,.
\end{array}
\right.
\end{equation}

One obtains a rather precise result for the range of parameters of interest
to us (see below) using the Boltzmann approximation:
\begin{equation}
\langle N_s-N_{\bar s}\rangle =
\gamma_s\left\{\int g_s\frac{d^3p}{(2\pi)^3}e^{-E/T}\right\}
\int_{R_{\rm f}} d^3r\left[\lambda_s e^{\frac V{3T}}
 - \lambda_s^{-1} e^{-\frac V{3T}}\right]\,.
\end{equation}
The Boltzmann limit allows also to verify the signs: the Coulomb
potential is negative for the negatively charged $s$-quarks with
the magnitude  of the charge, $1/3$, made explicit in the potential
terms in all expressions  above. It turns out that there is always
only one solution, with resulting $\lambda_s>1$. The magnitude of the
effect is quite significant: choosing $R_{\rm f}=8$\,fm, $T=140$\,MeV,
$m_s=200$\,MeV (value of $0.5 <\gamma_s<2$ is irrelevant) solution
of Eq.\,(\ref{Nsls}) for $Z_{\rm f}=150$ yields $\lambda_s=1.10$
(precisely: 1.0983, 1.10 corresponds to $R_{\rm f}=7.87$\,fm). This result
is consistent with one of the scenarios we
reported earlier  for Pb--Pb collisions \cite{Acta97}.
Thus we are reassured that the experimental data is
very likely consistent with deconfined quark source,
and hence a detailed verification of this hypothesis is needed.
The remarkable result we find is that experimental data is only
consistent with this value $\lambda_s=1.10\pm0.02$, see table \ref{fitpb}. 
Thus  as before for the
lighter system S--Au/W/Pb \cite{Let93,Raf91,LR98} we are finding that the
source of strange hadrons (up to Coulomb-asymmetry) is governed by
a symmetric, and thus presumably deconfined strange quark phase space.

\begin{table}[bt]
\caption{\label{fitpb}
Statistical parameters obtained seeking minimum of weighted least square 
difference with experimental results 
 shown in table \protect\ref{resultpb}.
Values of $\lambda_{\rm s}$ in approach D$_s$ is the result of 
strangeness conservation  constraint:
asterisk~$^*$  means a fixed (input) value, not a parameter.
In case  D$_t$ the temperature is fixed at the value obtained 
for  S--Au/W/Pb collisions, and in case D$_p$,
the pressure in the hadronic phase space is fixed at $82 \pm 6$ MeV/fm$^3$,
the value obtained in S--Au/W/Pb collisions.
In case F the four ratios with $\Omega$-data are studied.}
\vspace{-0.2cm}\begin{center}
\begin{tabular}{|l|ccccc|cc|}
\hline\hline
Fit&$T_{\rm f} [MeV]$& $\lambda_{\rm q}$&$\lambda_{\rm s}$&
$\gamma_{\rm s}$&$\gamma_{\rm q}$& $n_{\rm dof}$& $\chi^2/n_{\rm dof}$\\
\hline
                    A
                     &  147 $\pm$ 3
                 & 1.69 $\pm$ 0.03
                 &   1$^*$
                 &   1$^*$
                 &   1$^*$
                 &   8
                 &   16  \\

                    B
                     &  142 $\pm$ 3
                 & 1.70 $\pm$ 0.03
                 &   1.10 $\pm$ 0.02
                 &   1$^*$
                 &   1$^*$
                 &   7
                 &   13  \\

                    C
                     & 144 $\pm$ 4
                 & 1.62 $\pm$ 0.03
                 &   1.10 $\pm$ 0.02
                 &   0.63 $\pm$ 0.04
                 &   1$^*$
                 &   6
                 &   4  \\
\hline
                    D
                     &    134 $\pm$ 3
                 &   1.62 $\pm$ 0.03
                 &   1.10 $\pm$ 0.02
                 &   1.27 $\pm$ 0.08
                 &   1.84 $\pm$ 0.30
                 &   5
                 &    0.3       \\

                    D$_s$
                     &    133 $\pm$ 3
                 &   1.63 $\pm$ 0.03
                 &   1.09$^*$ $\pm$ 0.02
                 &   1.98 $\pm$ 0.12
                 &   2.75 $\pm$ 0.35
                 &   5
                 &    0.45       \\

                    D$_t$
                     &    143$^*$
                 &   1.61 $\pm$ 0.03
                 &   1.10 $\pm$ 0.02
                 &   0.74 $\pm$ 0.06
                 &   1.15 $\pm$ 0.18
                 &   5
                 &    0.8       \\

                    D$_p$
                     &    137 $\pm$ 4
                 &   1.62 $\pm$ 0.03
                 &   1.10 $\pm$ 0.02
                 &   0.88 $\pm$ 0.07
                 &   1.33 $\pm$ 0.10
                 &   5
                 &    0.5       \\
\hline
                    F
                     &    334 $\pm$ 18
                 &   1.61 $\pm$ 0.03
                 &   1.12 $\pm$ 0.02
                 &   0.09 $\pm$ 0.01
                 &   0.18 $\pm$ 0.02
                 &   8
                 &   2.4       \\
\hline\hline
\end{tabular}
\end{center}
\end{table}

Let us briefly explain how we obtain the statistical parameters 
shown in table \ref{fitpb}:
with $E_i=\sqrt{m_i^2+p^2}=\sqrt{m_i^2+p_\bot^2}\cosh y $
 we integrate over the transverse momentum range
as given by the experiment, see table \ref{resultpb}.
To obtain the relative strengths of centrally produced
particles we consider only central rapidity region $y\simeq 0$. 
 We  allow all hadronic resonances to disintegrate
in order to obtain the final relative multiplicity of `stable' particles
required to form the observed particle ratios.
We show results of four main models denoted A, B, C, D in top section of
 table \ref{fitpb}, arising  describing the  data  shown in the first 
4 columns of table \ref{resultpb}. In this group we
successively relax the chemical variables from their tacit values (= 1).
In each step the number of degrees of freedom  decreases by one, yet
as described by $\chi^2/n_{\rm dof}$ the confidence level becomes
progressively better, and is indeed of impressive 
quality, with $\chi^2/n_{\rm dof} = 0.3$  when all 4
chemical variables are allowed to vary. We thus conclude that it is 
necessary in description of the particle abundance data to allow 
non-equilibrium abundances of light and strange quarks.

\begin{figure}[tb]
\vspace*{0cm}
\centerline{\hspace*{1.5cm}
\psfig{width=12cm,figure=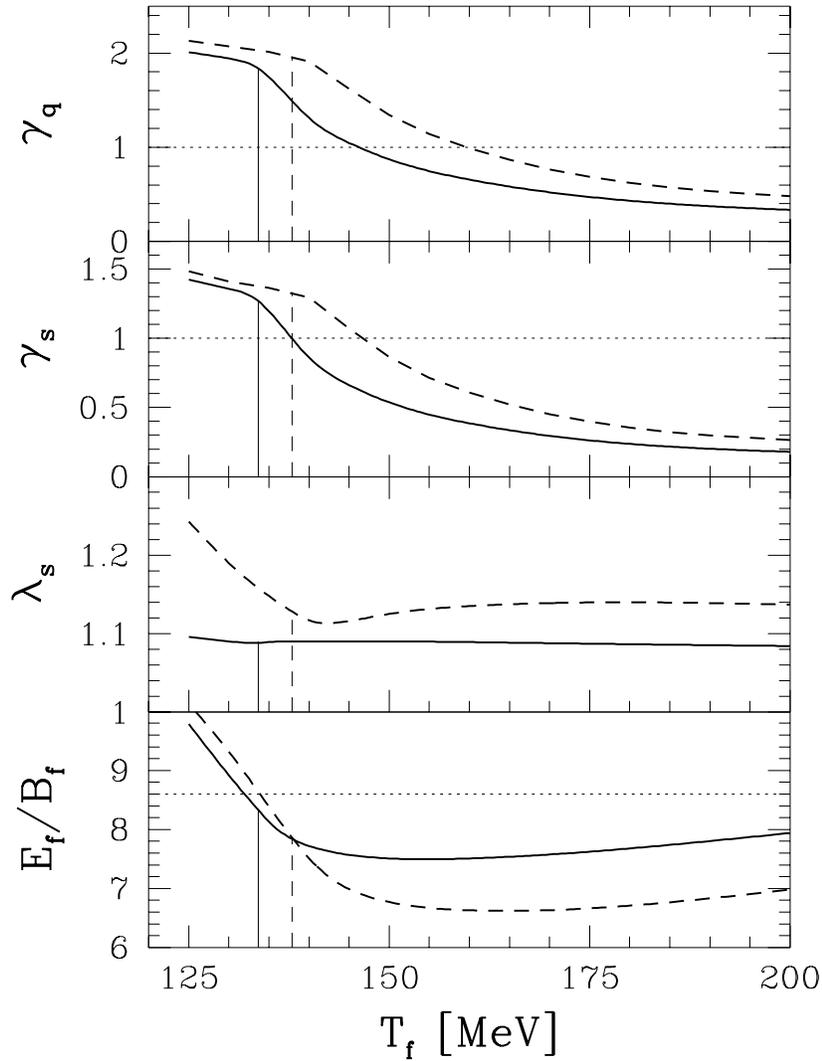}
}
\vspace*{-2.3cm}
\caption{Variation of $\gamma_{\rm q}$, $\gamma_{\rm s}$, $\lambda_{\rm s}$, 
 and $E_{\rm f}/B_{\rm f}$, 
as function of temperature $T_{\rm f}$, with all other parameters fixed by 
choosing best agreement with the experimental data.   
\label{TdepPb}
}
\end{figure}

This is confirmed by the results we show in Fig.\,\ref{TdepPb}, where 
for a given  $T_{\rm f}$, for case D (solid line, 
no $\Omega$ or strangeness conservation) and case D$_{\rm s}$
(dashed line, no $\Omega$, with   strangeness conservation), with 
all the other parameters obtained finding minimum of weighted 
least square theory-experiment difference at given 
temperature. The locations of the best $\chi^2/n_{\rm dof}$ are 
indicated by vertical lines. The top sections of the figure 
shows that both chemical non-equilibrium parameters 
$\gamma_{\rm s},\,\gamma_{\rm q}>1$ in a wide range of freeze-out temperatures, 
indeed the values are slightly higher for the  smaller 
$T_{\rm f}$ that are normally more favored on intuitive grounds (lower
freeze-out particle density). However,  freeze-out at $T_{\rm f}>145$\,MeV would 
allow values $\gamma_{\rm s},\,\gamma_{\rm q}<1$. It is reassuring that the 
analysis with matter flow \cite{LR99} even more clearly favors the high values of
$\gamma_{\rm s},\,\gamma_{\rm q}>1$ we found here. We note also the constancy
of the parameter $\lambda_{\rm s}$ (unconstrained  solid line result) which assures 
us that the Coulomb effect we described cannot be ignored. We also note 
the counter-intuitive result that the energy per baryon at freeze-out obtained 
from the properties of the hadronic phase space using the statistical parameters, 
is dropping as the freeze-out temperature increases in the region of interest here. 

\section{Hadronic particle abundances and phase space properties}\label{particles} 
Since we have the statistical parameters in hand, we can as alluded to above,
not only obtain the particle abundances within the kinematic cuts, shown in 
table \ref{resultpb}, but also study the physical properties of the hadronic system 
at freeze-out, shown in table \ref{tqpb}.

\begin{table}[tb]
\caption{\label{resultpb}
Particle ratios: Experimental results for Pb--Pb, references and kinematic
domain in first four columns, and different model results corresponding to
those shown  in table \protect\ref{fitpb} in the remaining columns;
Asterisk~$^*$ means the corresponding experimental result is not fitted.}
\vspace{-0.2cm}\small\begin{center}
\begin{tabular}{|lclc|llll|lll|l|}
\hline\hline
    Ratio& Ref. &cuts[GeV]         
&Exp. data&Fit A   &Fit B    &Fit C    &Fit D   &Fit D$_s$&Fit D$_t$&Fit D$_p$&Fit F\\
\hline
${\Xi}/{\Lambda}$                               &\protect\cite{WA97} &$p_{\bot}>0.7$&0.099 $\pm$ 0.008
&  0.130 &  0.138  &  0.093  &  0.095  &  0.098 &  0.095  &  0.093  &  0.107  \\
${\overline{\Xi}}/{\bar\Lambda}$                &\protect\cite{WA97} &$p_{\bot}>0.7$&0.203 $\pm$ 0.024
&  0.361 &  0.322  &  0.198  &  0.206  &  0.215 &  0.201  &  0.201  &  0.216  \\
${\bar\Lambda}/{\Lambda}$                       &\protect\cite{WA97} &$p_{\bot}>0.7$&0.124 $\pm$ 0.013
&  0.126 &  0.100  &  0.121  &  0.120  &  0.119 &  0.123  &  0.120  &  0.121  \\
${\overline{\Xi}}/{\Xi}$                        &\protect\cite{WA97} &$p_{\bot}>0.7$&0.255 $\pm$ 0.025
&  0.351 &  0.232  &  0.258  &  0.260  &  0.263 &  0.262  &  0.258  &  0.246  \\
$(\Xi+\bar{\Xi})\over(\Lambda+\bar{\Lambda})$   &\protect\cite{NA49} &$p_{\bot}>1.$ &0.13 $\pm$ 0.03
&  0.169 &  0.169  &  0.114  &  0.118  &  0.122 &  0.116  &  0.115  &  0.120  \\
K$^0_{\rm s}/\phi$                            &\protect\cite{NA49}&       &11.9 $\pm$ 1.5\ \
&  6.42  &  6.3    &  10.4   &  9.89   &  9.69  &  10.1   &  10.2   &  16.1   \\
K$^+$/K$^-$                                  &\protect\cite{NA49} &              &1.80  $\pm$ 0.10
&  2.27  &  1.96   &  1.75   &  1.76   &  1.73  &  1.73   &  1.77   &  1.62   \\
$p/{\bar p}$                                    &\protect\cite{NA49} &            &18.1 $\pm$4.\ \ \ \
&  20.5  &  22.0   &  17.1   &  17.3   &  17.9  &  16.6   &  17.3   &  16.7   \\
${\bar\Lambda}/{\bar p}$                        &\protect\cite{NA49} &              &3. $\pm$ 1.
&  2.97  &  3.02   &  2.91   &  2.68   &  3.45  &  2.91   &  2.96   &  0.65   \\
K$^0_{\rm s}$/B                               &\protect\cite{NA49} &              &0.183 $\pm$ 0.027
&  0.298 &  0.305  &  0.224  &  0.194  &  0.167 &  0.211  &  0.214  &  0.242  \\
${h^-}$/B                                       &\protect\cite{NA49} &              &1.83 $\pm $ 0.2\ \
&  1.30  &  1.47   &  1.59   &  1.80   &  1.86  &  1.55   &  1.72   &  1.27   \\
\hline
${\Omega}/{\Xi}$                                &\protect\cite{WA97} &$p_{\bot}>0.7$&0.192 $\pm$ 0.024
&  0.115$^*$ &  0.119$^*$  &  0.080$^*$  &  0.078$^*$  &  0.080$^*$ &  0.082$^*$  &  0.078$^*$  &  0.192  \\
${\overline{\Omega}}/{\overline{\Xi}}$          &\protect\cite{WA97} &$p_{\bot}>0.7$&0.27 $\pm$ 0.06
&  0.32$^*$  &  0.28$^*$   &  0.17$^*$   &  0.17$^*$   &  0.18$^*$  &  0.18$^*$   &  0.17$^*$   &  0.40   \\
${\overline{\Omega}}/{\Omega}$                  &\protect\cite{WA97} &$p_{\bot}>0.7$&0.38 $\pm$ 0.10
&  1.00$^*$  &  0.55$^*$   &  0.56$^*$   &  0.57$^*$   &  0.59$^*$  &  0.56$^*$   &  0.56$^*$   &  0.51   \\
$(\Omega+\overline{\Omega})\over(\Xi+\bar{\Xi})$&\protect\cite{WA97} &$p_{\bot}>0.7$&0.20 $\pm$ 0.03
&  0.17$^*$  &  0.15$^*$   &  0.10$^*$   &  0.10$^*$   &  0.10$^*$  &  0.10$^*$   &  0.10$^*$   &  0.23   \\
\hline\hline
\end{tabular}
\end{center}
\end{table}


We first observe, comparing the experimental data shown in left hand portion of table
\ref{resultpb} with the different theoretical results 
presented on the right hand side, that 
we are more successful in the description of particles above the horizontal line. 
Below, in  the bottom section of table \ref{resultpb} we group results comprising 
$\Omega$ and $\overline\Omega$-particles, which seem  not to follow 
the same systematics, a fact already reported with respect of
their spectral temperature by the WA97 collaboration \cite{WA97}.
A possible hypothesis is that  a good fraction of
these particles are made in processes that are different 
in nature than those leading to the other particle abundances, 
and hence we excluded  $\Omega$ and $\overline\Omega$-particles from all but
one of the approaches presented, denoted F. Moreover, if some strange particles 
hadronize separately, one cannot demand that the remaining particles balance
strangeness exactly and hence we did not in general enforce  strangeness
conservation, except in case D$_s$. That case is only slightly worse than  D, 
which implies that the  source for both $\Omega$ and $\overline\Omega$
is nearly symmetric with respect to abundance of strange and anti-strange quarks.
This is consistent with the observation that much of
the significant asymmetry in the ratio  $\overline\Omega$/$\Omega$
arises from the Coulomb effect we described above.
We note that model F shown in table \ref{fitpb}, including
the four $\Omega$-particle 
data points yields $\chi^2/n_{\rm dof}=2.4$ for $n_{\rm dof}=8$,
the mathematical confidence level is a few percent, it can 
safely be assumed that our approach is not adequately accounting for the 
production of $\Omega$-particle. 

We  address  extensively in our study the different constraints, 
as indicated by the subscript in all tables:
\begin{enumerate}\item
D$_s$ includes the requirement of strangeness conservation, {\it i.e.,} 
the hadronic phase space has to contain for the given statistical 
parameters as many $\bar s$- as $s$-quarks. This is most conveniently
accomplished by finding the value of $\lambda_s$ which balances 
strangeness in terms of the other 
parameters, and thus, though not fitted, the value shown in table
\ref{fitpb} displays an error, derived from the errors determined determining
the other statistical variables. We note that the phase space occupancies 
change drastically  between cases,  D and D$_s$, see table \ref{fitpb},
however the value 
$ \gamma_s/\gamma_q$ changes from 0.69 for  D to 0.72 for  D$_s$.
In actual numerical procedure we took advantage of this stability 
in $ \gamma_s/\gamma_q$-ratio, using it as a parameter. 
More generally, we note that all acceptable models shown in table \ref{fitpb}
yield $ \gamma_s/\gamma_q=0.68\pm0.05$, which is consistent with the result 
of model C for $\gamma_s$, where the tacit assumption $\gamma_q=1$ is made. 
\item
There is considerable reason to seek
a comparison of the Pb--Pb system with the analysis of S--Au/W/Pb reactions
which we reported earlier \cite{LR98}. Thus we consider 
model D$_t$ in which the freeze-out temperature 
is fixed at the value we found in S--Au/W/Pb reactions,
$T_{\rm f}=143$\,MeV. In model D$_p$,  the pressure  of the hadron phase
space is chosen at  the  value we found in S--Au/W/Pb reactions,
 $P=82\,\mbox{MeV/fm}^3$. As judged by   $\chi^2/n_{\rm dof}$ all D$_i$-models
are possible, and the resulting particle  multiplicities presented  in the 
columns of table \ref{resultpb} differ only in minute detail.
The two models D$_t$ and D$_p$ which test consistency with the smaller
S--Au/W/Pb reactions are well within the allowable error. This consistency
implies the possibility that the matter formed in these two very different
systems hadronize in a rather similar fashion, though collective surface flows
are very different. 
\end{enumerate}
To resolve if there is universal freeze-out we have to consider the
physical properties of the fireball. 
While the statistical parameters shown in table \ref{fitpb} can vary 
strongly from model to model, we  find that the implicitly determined
physical properties of the hadron source are more stable.  In
table \ref{tqpb} we show for the 8 models  along with their temperature the
specific energy and entropy content, and specific anti-strangeness content,
along with specific strangeness asymmetry, and finally pressure  evaluated by 
using the statistical parameters to characterize the hadronic particle 
phase space. We note that it is improper in general to refer to these properties 
as those of a `hadronic gas' formed in nuclear collisions, as the particles 
considered may be emitted in sequence, and thus there never is a 
stage corresponding to a hadron gas phase. However, in the 
event such a stage exists, we also evaluated (see last column in 
table \ref{tqpb}) the volume of the hadron 
gas source at chemical decoupling. In order to obtain this extensive
property, we used the net baryon number in the 
fireball being $\langle B-\bar B\rangle =372\pm10$, 
as stated in \cite{BGS98}. Note  that a spherical source corresponding
to the best model D would have a source radius 9.6 fm, which in turn can be
checked to be exactly in agreement with deconfined strangeness conservation
as described by Eq.\,(\ref{Nsls}), given the statistical parameters, and
$m_s\simeq200$\,MeV. Other interesting conclusions arising in view of
these results and shown in table  \ref{tqpb} are: the specific energy content
$E_{\rm f}/B$ is well within the expectations based on the collision energy
content per nucleon (8.6 GeV) and hence this result confirms firmly the
hypothesis that the energy stopping and baryon number stopping in the fireball
are very similar. The specific strangeness content of the Pb--Pb collision
 fireball is, by about 20\% smaller than S--Au/W/Pb result.

\begin{table}[tb]
\caption{\label{tqpb}
Physical properties of hadronic final state phase space
(specific energy, entropy, anti-strangeness, net strangeness,
pressure and volume) derived from statistical parameters shown in 
table \protect\ref{fitpb} which describes the Pb--Pb data in table
\protect\ref{resultpb}. Asterisk~$^*$ means fixed input. }
\vspace{-0.2cm}\begin{center}
\begin{tabular}{|l|cccccc|c|}
\hline\hline
Fit&$T_{\rm f}$ [MeV]& $E_{\rm f}/B$  & $S_{\rm f}/B$ & ${\bar s}_{\rm f}/B$ & 
$({\bar s}_{\rm f}-s_{\rm f})/B$ &$P_{\rm f}$ [GeV/fm$^3$] & $V_{\rm f}$ [fm$^3$] \\
\hline
                    A
                     &  147 $\pm$ 3
                 & 6.60 $\pm$ 0.40
                 &  37.0 $\pm$ 3
                 &  0.92 $\pm$ 0.05
                 &  0.29 $\pm$ 0.02
                 &  0.068 $\pm$ 0.005
                 &  6429 $\pm$ 500  \\
                    B
                     & 142 $\pm$ 3
                 & 7.13 $\pm$ 0.50
                 &  40.9 $\pm$ 3
                 &  1.02 $\pm$ 0.05
                 &  0.21 $\pm$ 0.02
                 &  0.053 $\pm$ 0.005
                 &  8994 $\pm$ 600  \\
                    C
                     &  144 $\pm$ 4
                 & 7.75 $\pm$ 0.50
                 & 41.7 $\pm$ 3
                 & 0.70 $\pm$ 0.05
                 & 0.14 $\pm$ 0.02
                 & 0.053 $\pm$ 0.005
                 & 10242 $\pm$ 800  \\
\hline
                    D
                 &   134 $\pm$ 3
                 &   8.33 $\pm$ 0.50
                 &   46.8 $\pm$ 3
                 &   0.61 $\pm$ 0.04
                 &   0.08 $\pm$ 0.01
                 &   0.185 $\pm$ 0.012
                 &   3619 $\pm$ 250  \\
                    D$_s$
                     &   133 $\pm$ 3
                 &  8.72 $\pm$ 0.50
                 &  48.1 $\pm$ 3
                 &  0.51 $\pm$ 0.04
                 &  0$^*$
                 &  0.687 $\pm$ 0.030
                 &  1134 $\pm$ 100  \\
                    D$_t$
                     &  143$^*$
                 & 7.63 $\pm$ 0.45
                 & 41.4 $\pm$ 3
                 & 0.68 $\pm$ 0.05
                 & 0.12 $\pm$ 0.01
                 &  0.072 $\pm$ 0.005
                 &  7517 $\pm$  500 \\
                    D$_p$
                     &  137 $\pm$ 4
                 & 8.05 $\pm$ 0.50
                 & 44.7 $\pm$ 3
                 & 0.67 $\pm$ 0.05
                 & 0.13 $\pm$ 0.01
                 & 0.082$^*$
                 & 7090 $\pm$ 500  \\
\hline
                    F
                     &  334 $\pm$ 18
                 & 9.79 $\pm$ 0.50
                 & 24.1 $\pm$ 2
                 & 0.78 $\pm$ 0.05
                 & 0.06 $\pm$ 0.01
                 & 1.64 $\pm$ 0.006
                 & 2303 $\pm$ 250  \\
\hline\hline
\end{tabular}
\end{center}
\end{table}


\section{Conclusions}\label{conc}
We have presented detailed analysis of hadron abundances observed
in central Pb--Pb interactions at 158 A GeV in terms of thermal equilibrium and
chemical non-equilibrium  phase space model of (strange) hadronic particles.
We assumed formation of a thermal dense matter fireball of a priori unknown structure, 
which explodes and disintegrates into the final state hadrons. We have presented 
several excellent descriptions of  all abundance
data which at present comprise 5 or more independent
 degrees of freedom, yielding a family of models
with acceptable confidence level. The physical statistical parameters
obtained  characterize a strange particle source which,
when allowing for Coulomb deformation of the strange and anti-strange quarks,
is exactly symmetric, as is natural {\it only} for a deconfined state. While
the  statistical parameters shown in table \ref{fitpb} can vary widely
there is no way to distinguish with naked eye the different 
approaches D, D$_s$, D$_t$, D$_p$ inspecting the particle abundances shown 
in  table \ref{resultpb}. It is important to take note that 
along with $\lambda_q=1.62\pm0.03,\lambda_s=1.10\pm 0.02$ there also 
is a stable value $ \gamma_s/\gamma_q=0.68\pm0.05$ under the different 
strategies one may follow to analyze the experimental data.
The chemical freeze-out temperature allowing for the systematic uncertainty 
seen in the acceptable group of models in table \ref{fitpb} 
is $T_{\rm f}=138\pm7$ MeV, and this implies that the freeze-out
baryochemical potential is $\mu_{\rm B}=199\pm3$\,MeV. The error
here is small, since the  best values  $T_{\rm f},\,\lambda_q$ are 
anti-correlated.

 Given these statistical parameters we have also evaluated the physical properties
 of the hadronic particle phase space, such as energy, entropy
and baryon number. The results  shown in table \ref{tqpb}
 describe the properties of the final state. These correspond nearly 
exactly to the initial state conditions,
confirming the consistency of our approach and validating the reaction
 picture  applied. This part of our analysis 
confirms that the reaction
proceeds by the way of the formation of a dense fireball comprising
highly excited hadronic matter.  In consistency with this we obtain
values of $\lambda_s$ which exactly match expectations for
strangeness balance in QGP, allowing for the Coulomb effect within
the particle source of the size $R_{\rm f}=9.6\pm2$\,fm. We compare 
conditions of the particle source for the two systems Pb--Pb and S--Au/W/Pb
and find that both can be seen as hadronizing in same physical conditions. 

We have compared the lighter system  S--Au/W/Pb \cite{LR98}, 
with the current  study of Pb--Pb, selecting comparable
freeze-out conditions (models D$_t$ and D$_p$). 
 We find, see table \ref{tqpb} and \cite{LR98} 
that the different physical properties of  the two hadronic source agree. 
This, along with the strange phase space symmetry and the Coulomb effect,
we believe that the sole possible interpretation the formation of a
deconfined phase in the initial stages of the collision, which subsequently
evolves and flows apart till it reaches the universal hadronization point, with 
many similar physical properties, independent of the collision system. 
System dependent will certainly be the surface collective 
velocity $\vec v_{\rm c}$ \cite{LR99}, 
however, our analysis was organized such that this vector field did not enter
here in a significant way.

We have begun,
using quark-gluon plasma equations of state which incorporate the
perturbative corrections and thermal masses, to study  detailed
scenarios of QGP formation and evolution that leads to the freeze-out
properties we obtained here. The important preliminary finding is that it
is possible to find QGP-fireballs that naturally lead to the  results 
obtained studying the experimental hadron abundance data, and thus the 
QGP hypothesis presented here is also consistent
with our current theoretical understanding 
of the QGP equations of state.

{\vspace{0.5cm}\noindent\it Acknowledgments:\\}
We thank E. Quercigh for interesting and stimulating discussions.\\
This work was supported in part
by a grant from the U.S. Department of Energy,  DE-FG03-95ER40937\,.\\
 LPTHE-Univ. Paris 6 et 7 is: Unit\'e mixte de Recherche du CNRS, UMR7589.



\begin{thebibliography}{99}\small
\setlength{\itemsep}{-.01cm}

\bibitem{FGL73}
H. Fritzsch, M. Gell-Mann and H. Leutwyler, {\it Phys. Lett.}
{\bf 47 B}, 365 (1973);\\
{H.D. Politzer}, {\it Phys. Rep.} {\bf 14},
129-180 (1974) (see p.\,154 and Refs.\,34--41).

\bibitem{HAG}
R. Hagedorn, Suppl. Nuovo Cimento
{\bf 2}, 147 (1965);
Carg\`ese lectures in Physics,
Vol.\,6, Gordon and Breach (New York 1977)
and references therein; \\
See also: J. Letessier,
H. Gutbrod and J. Rafelski, {\it Hot Hadronic Matter},
NATO-ASI series B346,  Plenum Press, New York 1995.

\bibitem{Acta97}
J. Rafelski, J. Letessier, and A. Tounsi,
{\it Acta Phys. Pol.}, B{\bf 28}, 2841 (1997);
 {\it Phys. Lett.} {\bf B 410}, (1997) 315.

\bibitem{LR98}
J. Letessier and J. Rafelski: 
``Chemical non-equilibrium and deconfinement in 200 A GeV Sulphur
   induced  reactions'',
 submitted to {\it Phys. Rev. C;} [hep-ph/9806386];\\
``Chemical non-equilibrium in High Energy Nuclear Collisions'',
submitted to {\it J. Phys. G} (proceeding of 
the Padova  --- Strangeness 1998 conference),
[hep-ph/9810332];\\
{J. Rafelski, J. Letessier and A. Tounsi},
{\it Acta Phys. Pol.} B {\bf 27}, 1035 (1996), and references therein.


\bibitem{HM96}
{J.W. Harris and B. M\"uller}, {\it Ann. Rev. Nucl. Science}
{\bf 46},  71 (1996), and references therein.

\bibitem{SAB}
J. Rafelski,  pp 282--324, 
GSI Report 81-6, Darmstadt, May 1981;
Proceedings of the Workshop on {\it Future Relativistic 
Heavy Ion  Experiments}, held at GSI, Darmstadt, 
Germany, October 7--10, 1980, R. Bock and R. Stock, Eds.,
(see in particular section 6, pp 316--320); see also:
 pp 619--632 in {\it New Flavor and Hadron Spectrosopy}, 
Ed. J. Tran Thanh Van (Editions Frontiers 1981), 
Proceedings of XVIth Rencontre de Moriond -- Second Session, 
Les Arcs, March 21--27, 1981; and:
{\it Nucl. Physics} A {\bf 374}, 489c (1982) --- Proceedings
of ICHEPNC held 6--10 July 1981 in Versailles, France;
{\it Phys. Rep.} {\bf 88}, 331 (1982); 
J. Rafelski and R. Hagedorn, in {\it Statistical Mechanics
of Quarks and Hadrons}, p.\,253, (North Holland, Amsterdam, 1981);
J. Rafelski and M. Danos, {\it Phys. Lett.} B {\bf 192}, 432 (1987).

\bibitem{Let93}
J. Rafelski, J. Letessier and A. Tounsi,
Dallas--ICHEP (1992) p.\,983 (QCD161:H51:1992), [hep-ph/9711350];\\
J. Letessier, A. Tounsi, U. Heinz, J. Sollfrank and J. Rafelski,
{\it Phys. Rev. Lett.} {\bf 70}, 3530 (1993).
{\it Phys.\ Rev.} D {\bf 51}, 3408 (1995).

\bibitem{entro}
{J. Letessier, A. Tounsi and J. Rafelski},
{\it Phys. Rev. }C {\bf 50}, 406 (1994);
{\it  Acta Phys. Pol.} A {\bf 85}, 699 (1994).


\bibitem{NA49}
H. Appelsh\"auser  {\it et al.}, NA49 Collaboration, ``$\Xi$ and $\overline\Xi$ 
production in 158 GeV/Nucleon Pb+Pb Collisions'', submitted to {\it Phys. Lett.} B
[nucl-ex/9810005];\\
G. Roland {\it et al.}, NA49 Collaboration, {\it Nucl. Phys.} A {\bf 638}, 91c (1998);\\
G.J. Odyniec,  {\it Nucl. Phys.} A {\bf 638}, 135c (1998);\\
F. P\"uhlhofer {\it et al.}, NA49 Collaboration, {\it Nucl. Phys.} A {\bf 638}, 431c (1998);\\
C. Bormann {\it et al.}, NA49 Collaboration,  {\it J. Phys.} G {\bf 23}, 1817 (1997);\\
G.J. Odyniec {\it et al.}, NA49 Collaboration, {\it J. Phys.} G {\bf 23}, 1827 (1997);\\
V. Friese{ \it et al.}, NA49 Collaboration, {\it J. Phys.} G {\bf 23}, 1837 (1997);\\
D. R\"ohrig {\it et al.}, NA49 Collaboration,``Recent results from NA49 experiment on 
Pb--Pb collisions at 158 GeV per nucleon'',
here see  figure 4, in proceedings of EPS-HEP Conference, Jerusalem, August 19-26, 1997;
available at http://www.cern.ch/hep97/abstract/tpa6.htm talk \#603;\\
P.G. Jones, for the NA49 Collaboration, {\it Nucl. Phys.} A {\bf 610}, 188c (1996).


\bibitem{BGS98}
F. Becattini, M. Gazdzicki and J. Sollfrank,
{\it Eur. Phys. J.} {C} {\bf 5}, 143 (1998).

\bibitem{WA97}
E. Andersen {\it et al.}, WA97-collaboration, 
{\it Phys.Lett.} B {\bf 433} 209 (1998);\\
I. Kralik {\it et al.}, WA97-collaboration, 
{\it Nucl.Phys} A{\bf 638} 115c (1998);\\
K. Safarik {\it et al.}, WA97-collaboration, 
{\it Nucl.Phys} A{\bf 630} 582 (1998);\\
A.K. Holme {\it et al.}, WA97 Collaboration, 
{\it J. Phys.} G {\bf 23}, 1851 (1997).

\bibitem{Kan97}
M. Kaneta {\it et al.}, NA44-collaboration, 
{\it J. Phys.} G {\bf 23}, 1865 (1997). 

\bibitem{HAGBB} H. Grote, R. Hagedorn and J. Ranft,
{\it Atlas of Particle Production Spectra},
(CERN-Service d'Information Scientifique, Geneva 1970).

\bibitem{BSWX96}
P. Braun-Munzinger, J. Stachel, J.P. Wessels and N. Xu,
{\it Phys. Lett.} B {\bf 365}, 1 (1996).

\bibitem{Raf91}
{J. Rafelski}, {\it Phys. Lett. }B {\bf 262}, 333 (1991);
{\it Nucl. Phys.} A {\bf 544}, 279c (1992).

\bibitem{SCRS93}
E. Suhonen, J. Cleymans, K. Redlich and H. Satz, in Proceedings of the 
International Europhysics Conference on High Energy Physics,
Marseille, France, 22-28 July 1993, 
 Marseille EPS HEP 1993, p.\,519 (QCD161:I48:1993), [hep-ph/9310345]. 

\bibitem{Hei94}
U. Heinz, {\it Nucl. Phys.} A {\bf 566}, 205 (1994)

\bibitem{Sol97}
J. Sollfrank, {\it J. Phys. } G {\bf 23}, 1903 (1997), 
and references therein.

\bibitem{Sae98}
Saeed-Uddin, {\it J. Phys.} G {\bf 24}, 779 (1998). 

\bibitem{GS98}
F. Grassi and O. Socolowski, Jr., 
{\it Phys. Rev. Lett.} {\bf 80}, 1170 (1998).


\bibitem{LR99}
``Hadronic particle chemical freeze-out with collective flow 
in 158 A GeV Pb-Pb collisions'',\\
J. Letessier and J. Rafelski, in preparation.


\bibitem{RM82}
{J. Rafelski and B. M\"uller}, {\it Phys. Rev. Lett.}
{\bf 48}, 1066 (1982); {\bf 56}, 2334E (1986);
{P.~Koch, B.~M\"uller and J.~Rafelski},
{\it Phys. Rep.} {\bf 142}, 167 (1986).

\bibitem{BZ82}
 T.S. Biro and  J. Zimanyi
{\it Phys. Lett.} B {\bf 113}, 6 (1982);
{\it Nucl. Phys.} A {\bf 395}, 525 (1983).

\bibitem{MR75}
B. Muller and J. Rafelski,
{\it Phys. Rev. Lett.} 34, 349 (1975).

\end{thebibliography}
\end{document}